\documentstyle[12pt,epsfig]{article}
\newcommand{\be}[1]{\begin{equation} \label{(#1)}}
\newcommand{\ee}{\end{equation}}
\newcommand{\ba}[1]{\begin{eqnarray} \label{(#1)}}
\newcommand{\ea}{\end{eqnarray}}

\begin{document}
\begin{center}
   {\Large\bf Relation between Lepton Flavor Violating Processes}\\[3mm]
Sergey Kovalenko and  Ivan Schmidt\\[1mm]
{\it Departamento de F\'\i sica, Universidad
T\'ecnica Federico Santa Mar\'\i a,}\\
{\it Casilla 110-V, Valpara\'\i so, Chile}
\end{center}
\bigskip

\begin{abstract}
We demonstrate the existence of model independent one-to-one correspondence relations
between different lepton flavor violating processes (LFV).
Applying the criterion of naturalness, based on the idea of ``custodial symmetry", we show that all the LFV
processes, independently of specific mechanisms behind them, but
with the same external leptons, have a priori comparable amplitudes
modulo their kinematics and involved form factors.
\end{abstract}
%
%
\bigskip
%

\bigskip

%
\bigskip
\bigskip

The discovery of neutrino oscillations has established the presence of neutrino masses
and mixing between neutrino species with different lepton flavors \cite{nu-oscill}.
This is the first observation of a lepton flavor violating (LFV) process, forbidden in
the SM.
Theoretically, the observed LFV cannot be isolated in the neutrino sector.
Indeed, it naturally extends to the sector of charged
leptons via the LFV charged current loop with virtual neutrinos, and then it should manifest itself in
the form of yet unobserved LFV processes with charged leptons, such as $\mu, \tau$ and meson LFV decays,
$\mu^--e^-$ nuclear conversion, etc.
The LFV in the charged lepton sector may also receive
contributions
directly from physics beyond the SM.
For this reason experimental and theoretical studies of LFV
charged lepton processes are expected to shed light on the
fundamental sources of LFV, offering one of the most direct ways
to observe physics beyond the SM.

In the present note we do not consider the possible origin of LFV.
Instead, we focus on the interplay between different LFV processes
in the sector of charged leptons. A central point of the present
study can be formulated as the question: if there is information
on some LFV process, say, an experimental upper limit for its rate
or its observation at a certain rate, then, what can be expected
for the rates of other LFV processes?
Gauge invariance and unitarity have already been applied in the
literature to relate the rates of some of the LFV processes
between each other \cite{Lit}.
The conclusions made in the present note, on the other hand,
derive from the absence of certain global ``custodial symmetry".
We will show that there exists a natural one-to-one correspondence
between the amplitudes of LFV processes with the same charged
lepton flavor structure, that is those processes which result in
the same change of the lepton flavors $L_i$. Our answer to the
above question is that the natural difference in the rates of the
LFV processes with the same lepton flavor structure should be
attributed to the difference in their kinematics and phase space,
whereas an additional strong suppression of some of these
processes with respect to others is unnatural and thus,
improbable.

Let us consider processes where LFV manifests itself via the
presence of two charged external leptons ($l_i, l_j$), and which
conserve baryon number $B$ and either conserve total lepton number
$\Delta L = 0$ (LFV) or violate it in two units $\Delta L = 2$
(LFNV). In addition to the two charged leptons these processes may
involve an arbitrary set of external particles satisfying certain
consistency conditions to be specified below.

For convenience we consider processes with the two leptons in the
final state and all the other participating particles in the
initial state. Nevertheless, as will be seen latter our
conclusions are valid for LFV and LFNV process with any
configuration of initial and final states.

Therefore, we focus on the processes:
$\Phi^{(0)}_k\rightarrow l_i^{-} l_j^{+}$ with $\Delta L = 0$, and
$\Phi^{(--)}_k\rightarrow l_i^{-} l_j^{-}$ with $\Delta L = -2$,
where $\Phi_k^{(0)}$ and $\Phi_k^{(--)}$ denote certain
color-singlet subsets of external particles with $B=0, L_i=0$ and
total electric charge $Q=0$ and $Q=-2$, respectively. The
difference between processes with the same lepton flavor structure
$l_i l_j$ resides in the difference of their external particle
subsets  $\Phi_k$. The possible sets of the SM particles in both
LFV and LFNV cases are of the following two types:
\begin{eqnarray}\label{set1}
&&\mbox{Type I:}\ \ \ \Phi_0^{(0)} \ \ = W^- W^+,\ \ \ \Phi_0^{(--)} = W^- W^-, \\
&&\mbox{Type II:}\ \  \Phi_k^{(0)} \ \  =  \left\{(\bar{u}u), (\bar{d}d), ... \right\}
\left\{(\bar{l}_i l_i), (W^- W^+), \gamma, Z, ... \right\},\\ \nonumber
&&\ \ \ \ \ \ \ \ \ \ \ \ \  \Phi_k^{(--)} =  (\bar{u} d)(\bar{u} d) \left\{(\bar{u}u), (\bar{d}d), ... \right\}
\left\{(\bar{l}_i l_i), (W^- W^+), \gamma, Z, ... \right\}.
\label{set2}
%
%
%
%
\end{eqnarray}
The particle sets of Type II for $\Phi_k^{(0)}$ are arbitrary
totally neutral collections of particles with total values $B=0,
L_i=0, Q=0$.  For the case of $\Phi_k^{(--)}$ the particle sets of
Type II must contain two pairs of anti-up and down quarks of any
generation to provide $Q=-2$. They may be accompanied  with any
neutral set of particles of the previous case. It is implied that
the quarks in $\Phi_k^{(0)}$ and $\Phi_k^{(--)}$ are arranged in
color singlet combinations forming mesons and/or baryons.

We write down the effective Lagrangians describing the LFV and LFNV processes in the following schematic form:
\begin{eqnarray}\label{eff-Lag-1}
{\cal L}_{_{LFV}} &=& \Phi_k^{(0)}\cdot \bar{l}_i \Gamma_{(k)ij} l_j  +
\Phi_k^{(0)}\cdot W^-_{\mu}{S}^{\mu\nu}_k W^+_{\nu} + \tilde{\cal L} + h.c. , \\
\label{eff-Lag-2}
{\cal L}_{_{LFNV}} &=& \Phi_k^{(--)} \cdot\bar{l}_i \Gamma^{\prime}_{(k)ij} l^c_j  +
\Phi_k^{(--)} \cdot W^+_{\mu}S^{\prime\mu\nu}_k W^+_{\nu}  + \tilde{\cal L}^{\prime} + h.c. .
\end{eqnarray}
The first effective operators in Eqs. (\ref{eff-Lag-1}) and (\ref{eff-Lag-2}) generate the amplitudes of
LFV $\Phi^{(0)}_k\rightarrow l^+_i l^-_j$ and LFNV $\Phi^{(--)}_k\rightarrow l^-_i l^-_j$ processes,
respectively, and may receive contributions
from the SM interactions via neutrino exchange
as well as from some physics beyond the SM. The vertex structure
$\Gamma_{(k)}$ depends on the concrete realization of this
operator. We also included two more terms in both Lagrangians
(\ref{eff-Lag-1}) and (\ref{eff-Lag-2}). The terms $\tilde{\cal
L}$ and $\tilde{\cal L}^{\prime}$ denote SM and any beyond the SM
interactions whose explicit form is irrelevant for our analysis.
We only state their possible presence. The second operators in
Eqs. (\ref{eff-Lag-1}) and (\ref{eff-Lag-2}) play a key role in
our analysis and are always induced by the SM interactions. This
is guaranteed by the structure of the field sets $\Phi_k^{(0)}$
and $\Phi_k^{(--)}$ given in Eqs. (\ref{set1}), (\ref{set2}).
The particular realization of these operators depends on the field
sets $\Phi_k^{(0)}$, $\Phi_k^{(--)}$  and determines the vertex
structures $S_k$ and $S^{\prime}_k$. The specific form of
$\Gamma_{(k)}, \Gamma^{\prime}_{(k)}$ and $S_k, S^{\prime}_k$ is
irrelevant for our subsequent reasoning. In what follows we denote
the first and second type of the effective operators in Eqs.
(\ref{eff-Lag-1}), (\ref{eff-Lag-2}) by $\hat{\Gamma}_{(k)ij},
\hat{\Gamma^{\prime}}_{(k)ij}$ and $\hat{S}_k,
\hat{S^{\prime}}_k$, respectively.

First, we observe that the operator $\hat{\Gamma}_{(k)ij}$ ($\hat{\Gamma^{\prime}}_{(k)ij}$)
not only contributes to the amplitude of the corresponding
LFV $\Phi^{(0)}_k\rightarrow l^+_i l^-_j$ (LFNV $\Phi^{(--)}_k\rightarrow l^-_i l^-_j$)
process but also to any other
LFV $\Phi^{(0)}_n\rightarrow l^+_i l^-_j$ (LFNV $\Phi^{(--)}_n\rightarrow l^-_i l^-_j$) process
with $n\neq k$ via the SM operators $\hat{S}_{k,n}$ ($\hat{S^{\prime}}_{k,n}$) as shown in Fig. 1.
%
Therefore, we can write down the following schematic formulas for
the amplitudes of the two different LFV processes
\begin{eqnarray}\label{Amplitudes1-1}
\Phi^{(0)}_k\rightarrow l^+_i l^-_j:\ \ \ {\cal A}_{(k)ij} &=& \langle \hat{\Gamma}_{(k)ij} \rangle +
\langle \hat{S}_k \hat{S}_n \hat{\Gamma}_{(n)ij} \rangle + \tilde{\cal A}_{(k)ij},\\
\label{Amplitudes1-2}
\Phi^{(0)}_{n\neq k}\rightarrow l^+_i l^-_j:\ \ \ {\cal A}_{(n)ij} &=& \langle \hat{\Gamma}_{(n)ij} \rangle +
\langle \hat{S}_n \hat{S}_k \hat{\Gamma}_{(k)ij} \rangle + \tilde{\cal A}_{(n)ij},
\end{eqnarray}
and also for the two different LFNV processes
\begin{eqnarray}\label{Amplitudes2-1}
\Phi^{(--)}_k\rightarrow l^-_i l^-_j:\ \ \ {\cal A}^{\prime}_{(k)ij} &=& \langle \hat{\Gamma^{\prime}}_{(k)ij} \rangle +
\langle \hat{S^{\prime}}_k \hat{S^{\prime}}_n \hat{\Gamma^{\prime}}_{(n)ij} \rangle + \tilde{{\cal A}^{\prime}}_{(k)ij},\\
\label{Amplitudes2-2}
\Phi^{(--)}_{n\neq k}\rightarrow l^-_i l^-_j:\ \ \
{\cal A}^{\prime}_{(n)ij} &=& \langle \hat{\Gamma^{\prime}}_{(n)ij} \rangle +
\langle \hat{S^{\prime}}_n \hat{S^{\prime}}_k \hat{\Gamma^{\prime}}_{(k)ij} \rangle + \tilde{{\cal A}^{\prime}}_{(n)ij},
\end{eqnarray}
where the objects between $\langle ...\rangle$ schematically
represent the contributions of the diagrams in Fig. 1 to the
matrix elements of these processes. The last terms in Eqs.
(\ref{Amplitudes1-1})-(\ref{Amplitudes2-2}) denote any other
contributions including those which originate from physics beyond
the SM.
From the presence of the second next-to-leading terms in the above relations
one may conclude that if the process $\Phi_k\rightarrow l_i l_j$
occurs at certain rate
then the process $\Phi_n\rightarrow l_i l_j$ should be also expected to occur at some non-zero rate
and visa versa.
Obviously, their
rates may differ significantly. One may expect that the natural suppression of one process
with respect to another be attributed to the difference in their kinematics, phase space
and, when the first term is eventually absent, also to the SM couplings involved
in the second terms in Eqs. (\ref{Amplitudes1-1})-(\ref{Amplitudes2-2}).
Going further, one may also admit the possibility of a strong accidental cancellation between the three terms
in the amplitudes of some processes. This could lead to an additional a priori unlimited suppression of these
processes with respect to the other ones. An extreme case of this type of suppression corresponds
to the vanishing of the amplitudes of some subset of the processes  ${\cal A}(\Phi_k\rightarrow l_i l_j) =0$, while
${\cal A}(\Phi_{n\neq k} \rightarrow l_i l_j) \neq 0$. This sort of suppression
is unnatural, since the above mentioned cancellation is, in
general, unstable with respect to radiative corrections. The only
stabilizer of such a cancellation is certain symmetry present in
the theory which protects ${\cal A}(\Phi_k\rightarrow l_i l_j) =0$
from radiative corrections to all orders. An adequate framework
for the analysis of the situation with hierarchical values of the
above amplitudes is given by the well known 't Hooft naturalness
rule \cite{tHooft} which, for the case of a hierarchy of two
observables ${\cal A}_1/{\cal A}_2\ll 1$, requires the presence of
a ``custodial symmetry" in the limit ${\cal A}_1/{\cal A}_2 =0$.
In order to study the possibility of hierarchy of the amplitudes
of the LFV and LFNV processes we examine whether there exists the
corresponding symmetry in the case of the Lagrangians
(\ref{eff-Lag-1}) and (\ref{eff-Lag-2}) including both the SM and
arbitrary beyond the SM interactions. This symmetry is associated
with the unitary transformations of the fields realizing
representations of a global group, which we denote by $G_{\eta}$.
Let the fields transform as
\begin{eqnarray}\label{group}
W^- \stackrel{G_\eta}{\longrightarrow} \eta_{_W} \cdot W^-,\ \
W^+ \stackrel{G_\eta}{\longrightarrow} \eta^{\dagger}_{_W} \cdot W^+,\ \
l_i \stackrel{G_\eta}{\longrightarrow} \eta_{_{li}}\cdot l_i, \ \ \
\Phi_k \stackrel{G_\eta}{\longrightarrow} \eta_k \cdot \Phi_k,\ \ \
\Phi^{\prime}_k \stackrel{G_\eta}{\longrightarrow} \eta^{\prime}_k \cdot \Phi^{\prime}_k,
\end{eqnarray}
where $\eta_a\cdot \eta_a^{\dagger} = 1$.
The effective operators $\hat{\Gamma}_{(k)ij}$ and $\hat{\Gamma^{\prime}}_{(k)ij}$,
generating the $\Phi^{(0)}_k\rightarrow l^+_i l^-_j$ and $\Phi^{(--)}_k\rightarrow l^-_i l^-_j$ processes,
transform as
\begin{eqnarray}\label{G1-eta}
\hat{\Gamma}_{(k)ij} \stackrel{G_\eta}{\longrightarrow} \eta_{(k)ij} \cdot \hat{\Gamma}_{(k)ij}, \ \ \ \mbox{with}\ \
\eta_{(k)ij} = \eta^{\dagger}_{_{li}}\cdot\eta_{_{lj}}\cdot \eta_k, \\
\label{G1-eta-prime}
\hat{\Gamma^{\prime}}_{(k)ij} \stackrel{G_\eta}{\longrightarrow} \eta^{\prime}_{(k )ij} \cdot \hat{\Gamma^{\prime}}_{(k)ij},
\ \ \ \mbox{with}\ \
\eta^{\prime}_{(k)ij} = \eta^{\dagger}_{_{li}}\cdot\eta^{\dagger}_{_{lj}}\cdot \eta^{\prime}_k
\end{eqnarray}
The group $G_{\eta}$ is not arbitrary and should be at least consistent with
the SM part of the Lagrangians (\ref{eff-Lag-1}) and (\ref{eff-Lag-2}),
represented by the operators $\hat{S}_k$ and $\hat{S^\prime}_k$. Their invariance
with respect to $G_{\eta}$ requires:
\begin{eqnarray}\label{constr}
\eta_k =1, \ \ \  (\eta^{\dagger}_{_W})^2\eta^{\prime}_k =1,
\end{eqnarray}
>From these constraints it follows that
\begin{eqnarray}\label{Gamma}
\eta_{(k)ij} = \eta_{(n)ij},\ \ \ \eta^{\prime}_{(k)ij} = \eta^{\prime}_{(n)ij} \ \ \ \mbox{for}\ \ \ \forall\  k,n\ .
\end{eqnarray}
The process $\Phi^{(0)}_k\rightarrow l^+_i l^-_j$
($\Phi^{(--)}_k\rightarrow l^-_i l^-_j$) is allowed by $G_{\eta}$
if $\eta_{(k)ij}=1$ ($\eta^{\prime}_{(k)ij}=1$) and forbidden if
$\eta_{(k)ij}\neq 1$  ($\eta^{\prime}_{(k)ij}\neq 1$). From
(\ref{Gamma}) it follows that if just one of the processes
$\Phi_k\rightarrow l_i l_j$ is allowed then all the ones with the
same external leptons $\Phi_{n\neq k}\rightarrow l_i l_j$ are
allowed and visa versa, and if at least one of them is forbidden
by $G_{\eta}$ all of them are also forbidden\footnote{Similar idea
was used in Refs. \cite{admis1,admis2} to prove one-to-one
correspondence between the amplitude of neutrinoless double beta
decay, Majorana neutrino \cite{admis1} and sneutrino \cite{admis2}
masses.}. Thus, we conclude that the ``custodial symmetry"
protecting a hierarchy of amplitudes of the LFV (LFNV) processes
with the same lepton flavor structure cannot be introduced in a
realistic theory.
Therefore, a priori, one may expect them to be of comparable order of magnitude.
Obviously, this conclusion does not take into account kinematics
and hadronic structure factors which may result in a significant,
but theoretically controllable, difference between the rates of
different processes.

As we already pointed out the above result is valid in the presence of any physics beyond the SM.
It is valid not only for LFV (LFNV) processes of the type
$\Phi^{(0)}_k\rightarrow l^+_i l^-_j$ ($\Phi^{(--)}_k\rightarrow
l^-_i l^-_j$) but for any process with arbitrary configurations of
final and initial states like $\Phi_{k1}\rightarrow \Phi_{k2}
l^+_i l^-_j$ ($\Phi^{\prime}_{k1}\rightarrow \Phi^{\prime}_{k2}
l^-_i l^-_j$), $\Phi_{k1} l^-_i \rightarrow \Phi_{k2}  l^-_j$
($\Phi^{\prime}_{k1} l^+_i \rightarrow \Phi^{\prime}_{k2}
l^-_j$). This is because our arguments are based on the analysis
of the effective operators $\hat{\Gamma}_{(k)ij}$, which generate
all these processes. The most typical examples of them are: meson
decays, $M^0\rightarrow l^+_i l^-_j$, $M_1^+\rightarrow M_2^-
l^+_i l^+_j$; neutrinoless nuclear double beta decay,
$(A,Z)\rightarrow (A,Z+2) l^-_i l^-_j$; muon nuclear conversion,
$\mu^- (A,Z)\rightarrow  e^{-} (A,Z)$; $\mu\rightarrow e \gamma$,
etc.

Let us illustrate the above general conclusion by considering the
concrete case of the two processes: (a) $\pi^0\rightarrow l^+_1
l^-_2$ and (b) $W^+W^-\rightarrow l^+_1 l^-_2$. They may receive
contributions from the effective operators of a different high
scale origin. In particular, the operators
$O^{(6)}=(\bar{q}\gamma_5 q\ \cdot \bar{l}_1 \gamma_5
l_2)/\Lambda^2$ and
$O^{(5)}=(W^+_{\mu}W^-_{\nu}\ \cdot  \bar{l}_1 \sigma^{\mu\nu}
l_2)/\Lambda$,  corresponding to the dominant tree level
contributions to (a) and (b) processes, respectively, have
different physical dimensions, dim=6 and dim=5, and originate from
different underlying renormalizable interactions after the
integrating out different heavy states. For this reason the rates
of the two processes may look totally unrelated and the situation
when one process is strongly suppressed with respect to the other
does not seem theoretically inconsistent.
%
%
However, at loop level there are cross contributions of  $O^{(6)}$
and $O^{(5)}$ operators: $O^{(6)}$ contributes to
$\pi^0\rightarrow l^+_1 l^-_2$ and $O^{(5)}$ to $W^+W^-\rightarrow
l^+_1 l^-_2$ in the way shown in Fig. 1. Now again one can imagine
the situation when in certain underlying model the common
contribution of both operators at tree and loop levels (including
the possible contributions of some other operators) to one of
these processes, say  to the process (a), is,  due to
self-cancellation, strongly suppressed or even vanishes resulting
in zero amplitude of (a), while this does not happen in process
(b), leaving its amplitude at a sizable level. The general
conclusion made in the present paper excludes such a situation and
requires that if one of these processes occurs with some non-zero
amplitude then the other also has  a  non-zero  amplitude. Moreover, any
hierarchy in the magnitude of these amplitudes,
${\cal A}(\pi^0\rightarrow l^+_1 l^-_2) \gg {\cal A}(W^+W^-\rightarrow
l^+_1 l^-_2)$ or ${\cal A}(\pi^0\rightarrow l^+_1 l^-_2) \ll
{\cal A}(W^+W^-\rightarrow l^+_1 l^-_2)$, is shown to be unnatural.
Of course, the notion of naturalness is, as always, understood as
a comparability in order of magnitude.

In conclusion, we have shown that the 't Hooft naturalness rule \cite{tHooft} implies that
different LFV(LFNV) processes with the same external leptons all have comparable
amplitudes modulo their kinematics and form factors of the participating hadrons.
This model independent conclusion is valid in the presence of any
physics beyond the Standard Model and should be helpful for the theoretical evaluation of 
the observability of the LFV processes in the future experiments.

%

\vskip15mm
\centerline{\bf Acknowledgments}

This work was supported by the FONDECYT projects 1030244, 1030355
\bigskip

\vspace*{1cm}

\begin{figure}\label{interplay}
\begin{center}
\epsfig{file=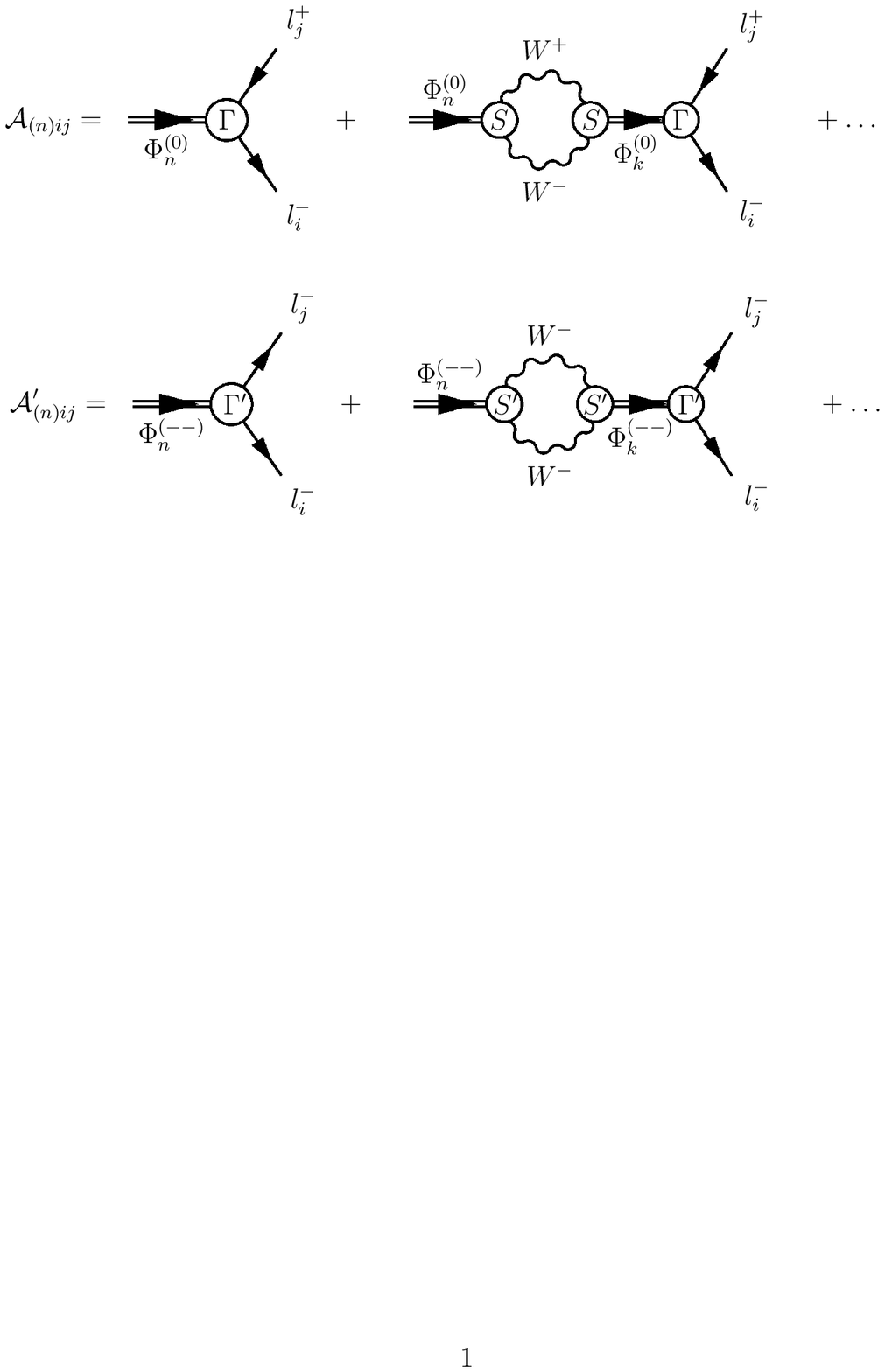, scale=1.}\\[-120mm]
\caption{Structure of the LFV and LFNV amplitudes.}
\end{center}
\end{figure}

\end{document}